\documentclass[9pt,conference]{IEEEtran}

\usepackage[preprint]{waspaa25}

\usepackage{bm} 

\title{Learning Robust Spatial Representations from Binaural Audio through Feature Distillation}

\name{Holger Severin Bovbjerg$^{1}$,
      Jan Østergaard$^{1}$,
      Jesper Jensen$^{1, 2}$,
      Shinji Watanabe$^{3}$,
      Zheng-Hua Tan$^{1}$\thanks{This work was partially funded by the William Demant Foundation.}\thanks{To appear in Proc. WASPAA 2025, October 12-15, 2025, Tahoe, US}.} %
\address{$^{1}$Aalborg University, Aalborg, Denmark \;
$^{2}$Eriksholm Research Centre, Snekkersten, Denmark\\
$^{3}$Carnegie Mellon University, Pittsburgh, USA
}

\usepackage[font=footnotesize,labelfont=bf]{caption}
\usepackage{csquotes}
\usepackage{cleveref}

\usepackage{siunitx}
\DeclareSIUnit\byte{B}

\usepackage{booktabs}
\usepackage{multirow}
\usepackage{stfloats}

\newcommand{\abs}[1]{\left\lvert#1\right\rvert}
\newcommand{\norm}[1]{\left\lVert#1\right\rVert}
\newcommand{\complexconjugate}[1]{%
  \overline{#1}%
}
\DeclareMathOperator*{\argmax}{arg\,max}

\def\realnumbers{\mathbb{R}}
\def\imagnumbers{\mathbb{I}}

\usepackage{hyperref}

\begin{document}

\maketitle

\begin{abstract}
Recently, deep representation learning has shown strong performance in multiple audio tasks. However, its use for learning spatial representations from multichannel audio is underexplored.
We investigate the use of a pretraining stage based on feature distillation to learn a robust spatial representation of binaural speech without the need for data labels.
In this framework, spatial features are computed from clean binaural speech samples to form prediction labels. 
These clean features are then predicted from corresponding augmented speech using a neural network.
After pretraining, we throw away the spatial feature predictor and use the learned encoder weights to initialize a DoA estimation model which we fine-tune for DoA estimation.
Our experiments demonstrate that the pretrained models show improved performance in noisy and reverberant environments after fine-tuning for direction-of-arrival estimation, when compared to fully supervised models and classic signal processing methods.
\end{abstract}


\section{Introduction}
Through evolution, animals, including humans, have developed the ability to extract spatial information from audio signals, allowing them to determine the direction-of-arrival (DoA) of a sound source, even in adverse conditions with noise and reverberation. 
This ability is based on differences in interaural time difference (ITD) and level difference (ILD), along with cues from the shapes of the head and ears \cite{Grothe_mammal_hearing}. 
Many signal processing algorithms try to replicate this ability for applications such as hearing aids, robots, and teleconferencing.
These methods generally try to extract the interaural phase difference (IPD) which is directly related to the ITD.
Popular algiorithms include the generalized cross-correlation (GCC) and its phase transform version (GCC-PHAT) \cite{Knapp_GCC_for_time_delay_estimation, Omologo_CPS_event_location, Benesty_microphone_array_SP, Blandin_GCCPHAT_clustering, Xiao_GCC_learning_based_DoA}, steered response power (SRP) \cite{Grinstein_SRP_tutorial} and multiple signal classification (MUSIC) \cite{Schmidt_MUSIC}.
These classic methods usually perform well in simple environments; however, their performance deteriorates significantly in a noisy and reverberant environment \cite{Xiao_GCC_learning_based_DoA}.

Data-driven models like Gaussian mixture models (GMMs) \cite{Fuchs_monaural_sound_localization} and deep neural networks (DNNs) \cite{Ma_DNN_robust_localization, yalta_dnn_doa, subramanian_dnn_multisource_localization, Abdelrahman_ssl_speech_review} have also been proposed for DoA estimation.
DNNs, in particular, have shown to excel in noisy and reverberant environments \cite{Ma_DNN_robust_localization, Xiao_GCC_learning_based_DoA}. 
In the DNN setup, one or more hand-crafted spatial features, such as GCC-PHAT or IPD, are usually passed to the model as input features.
Alternatively, some DNN-based methods use short-time Fourier transform (STFT) features directly, instead of relying on hand-crafted spatial features.

The idea of enhancing hand-crafted spatial features has also been explored \cite{Pak_phase_difference_enhancement, Cheng_steering_vector_enhancement}.
In \cite{Pak_phase_difference_enhancement}, the authors propose to enhance IPD features, using oracle IPDs as the target. This approach was found to perform better than the direct estimation of DoAs from IPDs.
Most recently, IPDNet \cite{Wang_IPDNet} was proposed, which predicts narrowband direct path IPDs from STFT input features. 
Here, the target is formed by theoretically deriving narrowband IPDs for the DoA label, instead of directly predicting the DoA.

The area of representation learning has shown that DNNs can learn highly performative representations of audio data.
Especially, self-supervised learning (SSL) has gained traction in audio signal processing, as it enables learning from unlabelled data \cite{Huang_masked_audio_MAE,unsupervised_AR_model_for_speech, baevski_wav2vec2, HuBERT_paper, Chen_WavLMLS, Chen_BEATS_SSL}. 
SSL describes a paradigm in which pseudo-labels are automatically derived from unlabelled data. The pseudo-lables are then used for training a neural network on a pretext task, with the goal of learning a good data representation.
Examples of pretext tasks include predicting future values \cite{unsupervised_AR_model_for_speech}, masked prediction \cite{Huang_masked_audio_MAE, HuBERT_paper, baevski_wav2vec2, Chen_BEATS_SSL} and contrastive learning \cite{baevski_wav2vec2, oord_cpc}.

Although much effort has been put towards learning robust audio representations \cite{Chen_WavLMLS, Chen_XEUS, Zhu_noise_robust_ssl_asr}, the focus has largely been on single-channel audio, and spatial representation learning remains underexplored. 
Some self-supervised/unsupervised approaches have been proposed \cite{Munakata_joint_sep_and_loc_FSC, Wang_UNSSOR}, although they mainly focus on speech separation, rather than spatial learning, and do not operate on binaural audio. 
Recently, SSLSAR \cite{Yang_SSLSAR} was proposed, using cross-channel reconstruction as the pretext task in binaural setups. 
However, this work does not address noisy, reverberant data or causal frame-level prediction, essential for many applications, such as DoA estimation in hearing aids.



This paper investigates the use of spatial features as spatial representation learning targets and proposes a pretraining framework catered towards learning robust causal frame-level spatial features without the need for DoA labels by predicting spatial features computed from clean speech from noisy speech, similar to WavLM \cite{Chen_WavLMLS}.
To evaluate the learned representation, we fine-tune the pretrained models to predict the direction of arrival of a speech source.

Due to the lack of readily available large binaural data sets with annotated direction of arrivals, we create a benchmark data set based on Librispeech \cite{librispeech} and LibriLight \cite{librilight} data, which is binauralized using recorded head-related transfer functions (HRTF) from the ARI HRTF database \cite{ari_hrtf}.
We systematically evaluate the DoA estimation performance in different noisy environments and at various noise levels to assess the robustness of the learned representations. 

The key contributions of this paper are:
\begin{itemize}
    \item A novel learning approach for learning robust causal spatial features from binaural speech without the use of spatial annotations.
    \item A comparison of different spatial features for use as spatial representation learning targets.
    \item An extensive evaluation of various binaural DoA estimation approaches in clean and noisy conditions, demonstrating the effectiveness of our pretraining framework for learning a robust spatial representation.
\end{itemize}
    

The code and data set\footnote{\url{https://github.com/HolgerBovbjerg/spatial_feature_distillation}} used for model training and generation of the results presented in this paper are publicly available.

\section{Direction-of-arrival Estimation}\label{sec:doa_estimation}
For a multichannel audio signal, the observed signal can be described by \eqref{eq:signal_model}:
\begin{equation}
    x^m[t] = \sum_{i=0}^{N_\mathrm{s}-1} (\bm{h}^{i, m} \ast \bm{s}^{i})[t] + v[t]^m, \label{eq:signal_model}
\end{equation}
where $\bm{x}^m[t]$ represents the $t$'th sample of the observed signal at microphone channel $m$, $N_\mathrm{s}$ is the number of sources, $\bm{h}^{i, m}$ is the impulse response related to source $i$ and microphone $m$, $\bm{s}^{i}$ is the $i$'th source signal and $v^m[t]$ represents additive noise.

Signal analysis is usually performed in a framewise manner in the frequency domain by applying STFT to the observed signal \cite{Grinstein_SRP_tutorial}. 
This results in STFT frames: 
\begin{equation}
    \bm{X}^m[n] = \mathrm{STFT}(\bm{x}^m)[n], \label{eq:stft}
\end{equation}
where $\mathbf{X}^m[n]$ is the sequence of STFT frame representations of the framed observed signal $\bm{x}^m$ for the $n$'th frame.

The relevant information for estimating the direction of arrival of a source signal $\bm{s}$, is contained in the ITD, ILD and other cues arrising due to $\bm{h}$ in \eqref{eq:signal_model}.

\subsection{Model-based DoA estimation}
Classic model-based methods generally assume that a single source signal is dominant, and rely on estimating the ITD which in combination with the array geometry can be used to estimate the DoA.
Assuming a binaural microphone setup, the DoA can be estimated from the ITD as:
\begin{equation}
    \phi[n] = \arcsin\left(\frac{c \cdot \mathrm{ITD}^{m_1, m_2}[n]}{\norm{\mathbf{d}^{m_1, m_2}}}\right),\label{eq:gcc_argmax_angle}
\end{equation}
where $\mathbf{d}^\mathrm{m_1, m_2}$ is a vector representing the distance between the spatial positions of microphones $m_1$ and $m_2$, $c$ is the speed of sound, and $\bm{\phi}[n]$ is the estimated DoA for the $n$th frame relative to the normal vector of $\mathbf{d}^\mathrm{m_1, m_2}$.

The ITD is directly related to the IPD which can be found by computing the cross-power spectrum (CPS):
\begin{align}
    \mathrm{CPS}^{m_1, m_2}[n, f] &= X^{m_1}[n, f] \complexconjugate{X}^{m_2}[n, f],  \label{eq:cps}
\end{align} 
where $\mathrm{CPS}^{m_1, m_2}[n, f]$ denotes the CPS coefficient of the $n$'th frame at frequency index $f$ between microphones $m_1$ and $m_2$, and $\complexconjugate{X}$ denotes the complex conjugate.

The IPD is found as the phase of the CPS:
\begin{equation}
    \mathrm{IPD}[n, f] = \arg(\mathrm{CPS}^{m_1, m_2}[n, f]) \label{eq:ipd}
\end{equation}

The ITD can then be estimated from the IPD as:
\begin{equation}
\mathrm{ITD}^{m_1, m_2}[n, f] = \frac{\mathrm{IPD}^{m_1, m_2}[n, f]}{\omega[f]}, \label{eq:ipd_to_itd}
\end{equation}
where $\omega[f]$ is the angular frequency corresponding to frequency bin $f$.

Another way to estimate the ITD, is to transform the CPS to time domain by applying an inverse discrete Fourier transform (IDFT) to obtain the generalized cross-correlation (GCC) \cite{Knapp_GCC_for_time_delay_estimation, Omologo_CPS_event_location, Grinstein_SRP_tutorial, Blandin_GCCPHAT_clustering}:
\begin{equation}
    \mathrm{GCC}^{m_1, m_2}[n, \tau] = \mathrm{IDFT}(\alpha^{m_1, m_2} \mathbf{CPS}^{m_1, m_2}[n])[\tau], \label{eq:gcc_phat}
\end{equation}
where $\tau$ denotes a specific delay between $m_1$ and $m_2$ in number of samples, and $\alpha$ is an optional frequency-domain weighting function.

The weighting function $\alpha$ is commonly chosen as the \textit{phase transform} (PHAT), which has been shown to be more robust to noise. 
PHAT-weighting whitens the spectrum by normalizing each CPS frequency component in \eqref{eq:cps} by its magnitude, such that only the phase information is kept:
\begin{equation}
    \alpha_\mathrm{PHAT}^{m_1, m_2}[n, f] = \abs{\mathrm{CPS}^{m_1, m_2}[n, f]}^{-1}, \label{eq:cps_phat}
\end{equation}
where $\alpha_\mathrm{PHAT}^{m_1, m_2}$ is the PHAT-weighting.

To estimate the ITD, one can simply find the value of $\tau$ that maximizes \eqref{eq:gcc_phat}, and divide by the sampling frequency to convert it to seconds:
\begin{equation}
\mathrm{ITD}^{m_1, m_2}[n] = \frac{\argmax_\tau{(\mathrm{GCC}}_\mathrm{PHAT}^{m_1, m_2}[n, \tau])}{F_\mathrm{s}},\label{eq:gcc_argmax_tdoa}
\end{equation}
where $F_\mathrm{s}$ is the sampling frequency.







\subsection{DNN-based DoA estimation} \label{subsec:doa_dnn}
While model-based features like GCC and IPD have been shown to be good estimators of the DoA, they are prone to noise and reverberation.
Some studies have investigated using GCC or IPD as the input features to a DNN to improve the DoA prediction performance \cite{Kowalk_DNN_signal_informed_DoA_w_GCCPHAT, Kowalk_DoA_gemoetry_aware_dnn}, combining the model-based and a data-driven approaches.

Another approach is to rely on the DNN to learn relevant spatial information, such as channel correlations, instead of using hand-crafted spatial features.
A common setup is to use STFT features as input to the DNN \cite{Goli_STFT_DNN_localization, Yang_SSLSAR}. 
Here, STFT features are first computed for each channel, as in \eqref{eq:stft}. 
The real and imaginary parts are then concatenated for each channel to form a real-valued input feature for each channel.
Finally, the channelwise features are concatenated to form a single feature vector, such that
\begin{align}
    \mathbf{\hat{X}}^m[n] &= \left[ \mathbf{X}^m_\realnumbers[n], \mathbf{X}^m_\imagnumbers[n] \right] \\
    \mathbf{\hat{X}}[n] &= \left[ \mathbf{\hat{X}}^{m_1}[n], \mathbf{\hat{X}}^{m_2}[n] \right] \label{eq:stft_real_imag}
\end{align}
where $\mathbf{\hat{X}}[n]$ denotes a single feature vector containing real and imaginary parts of the STFT, $\mathbf{X}^m_\realnumbers$ and $\mathbf{X}^m_\imagnumbers$, from microphone channels $m_1$ and $m_2$ corresponding to the $n$'th frame.

Using \eqref{eq:stft_real_imag} as input, allows the DNN more freedom to potentially learn a spatial representation that is more robust to noise and reverberation than GCC and IPD features.
However, the downside is that achieving good performance requires vast amounts of labelled training data which can be difficult to obtain.

The DoA estimation problem can be posed as a linear regression task, e.g., using a Haversine loss \cite{Tang_RegressionAC} between the predicted and true DoA. 
However, most DoA estimation DNNs are trained using cross-entropy, by quantizing the possible range of DoAs to a fixed set of $k$ directions, $\{\theta_1, \theta_2, \dots, \theta_k\}$, assigning the class label as the angle closest to the true DoA. 
In this work, we follow the classification approach using a cross-entropy loss.


\section{Feature Distillation Framework}

In the proposed setup, a neural network, consisting of an encoder and a feature predictor, is first trained to predict spatial features computed from clean speech, based on noisy input speech. 
Whereas feature enhancement methods \cite{Pak_phase_difference_enhancement, Cheng_steering_vector_enhancement, Wang_IPDNet} use the predicted features to directly predict DoAs, we discard the feature predictor and instead use the learned encoder representation for downstream DoA prediction, similar to many SSL methods \cite{HuBERT_paper, Chen_WavLMLS, Yang_SSLSAR}.

As described in \Cref{sec:doa_estimation}, classic spatial features are effective for DoA estimation in relatively clean conditions. 
DNN based approaches can improve estimation in noisy and revereberant environments, however, with the need for training data with DoA labels. 

\begin{figure}[tb]
    \centering
    \includegraphics[width=0.8\linewidth]{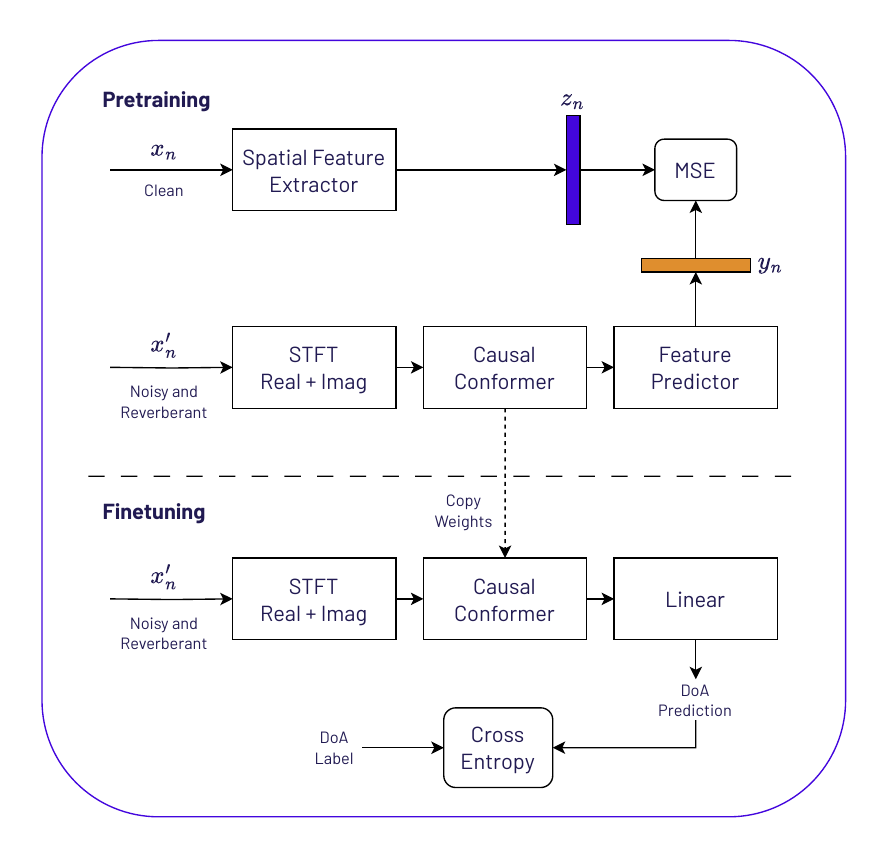}
    \caption{Spatial Feature Distillation framework overview.}
    \label{fig:neural_gcc}
\end{figure}

Inspired by the spatial feature enhancement methods, we propose the use of spatial features as representation learning targets in a feature distillation setup.
We hypothesize that classic spatial features serve as effective training targets for learning robust spatial features, removing the need for DoA labels. 
Specifically, we propose a pretraining task where a DNN predicts spatial features, computed from a clean microphone signal, from a noisy and reverberant input, similar to some noise-robust SSL methods \cite{Chen_WavLMLS, Chen_XEUS, Zhu_noise_robust_ssl_asr}. 
We refer to this framework as \textit{Spatial Feature Distillation} (SFD). 
An illustration of the proposed framework is shown in \Cref{fig:neural_gcc}.

As depicted in \Cref{fig:neural_gcc}, spatial features are computed from a clean non-reverberant binaural signal, as in \eqref{eq:gcc_phat}, yielding target vectors:
\begin{equation}
    \mathbf{z}[n] = \mathrm{SFE}(X^{\prime\,m_1}[n], X^{\prime\,m_2}[n]), \label{eq:neural_gcc_target}
\end{equation}
where $\mathbf{z}[n]$, is the target vector for the $n$'th frame and $\mathbf{SFE}$ is a spatial feature extractor. Here we use either GCC/GCC-PHAT, as described in \eqref{eq:gcc_phat}, or IPD from \eqref{eq:ipd} concatenated with the + ILD.

In the other branch, a corresponding noisy and reverberant signal is passed through an STFT feature extractor, as in \eqref{eq:stft_real_imag}. 
The STFT features are then encoded by a causal Conformer encoder \cite{Gulati_conformer_interspeech} followed by a prediction module, producing predicted spatial features:
\begin{equation}
     \mathbf{y}[n] = \operatorname{Linear}\left( \operatorname{Conformer}\left( \mathbf{\hat{X}^{\prime}}[n] \right) \right), \label{eq:neural_gcc_pred}
\end{equation}
where $\mathbf{y}[n]$ is the prediction for the $n$'th frame , and $\mathbf{\hat{X}}[n]$ denote concatenated STFT features as in \eqref{eq:stft_real_imag}.

The model is trained to minimize the MSE between $\mathbf{y}[n]$ from \eqref{eq:neural_gcc_pred} and $\mathbf{z}[n]$ from \eqref{eq:neural_gcc_target} such that:
\begin{equation}
    \mathcal{L}_\mathrm{MSE} = \frac{1}{N}\sum_{n=0}^{N-1}{\norm{\mathbf{y}[n] - \mathbf{z}[n]}_2^2}, \label{eq:mse_loss}
\end{equation}
where $\mathcal{L}$ denotes the loss, $n$ is the frame index and $N$ is the number of frames per sample.

After pretraining, we use the pretrained model weights to initialize a DoA prediction model, consisting of a conformer encoder with a linear classification head, as shown in \Cref{fig:neural_gcc}.

\section{Experimental Setup}
In this study, we consider a scenario with a single static speech source in a noisy environment and a binaural microphone setup. 
The method can be extended to incorporate more microphones and accommodate more complex environments; however, we will leave this for future work. 

\subsection{BinauralLibrispeech Dataset} \label{subsec:binaural_dataset}
Currently, there are no large publicly available binaural speech datasets, especially those with source direction and microphone geometry annotations. 
Therefore, we simulate binaural data for our experiments, similar to \cite{Barker_Clarity_challenge}. 
While this may limit real-world generalization, our focus is on exploring learning a spatial audio representation from large unlabelled binaural datasets, enabling its use when available - whether simulated or real.

Our binaural dataset is based on LibriSpeech \cite{librispeech} and LibriLight \cite{librilight} utterances which we convolve the left and right HRTFs from the ARI HRTF dataset \cite{ari_hrtf} to generate binaural signals. 
To add reverb we use simulated room impulse responses from SLR28 \cite{Ko_RIR_data_augmentation}.

To prevent HRTF leakage, we split the ARI HRTF dataset into training, validation, and test sets, with an 80:10:10 split of subjects.
HRTFs are sourced from over 250 subjects, recorded with a spatial resolution of \SI{5} {\degree}. 
The training data is based on LibriLight 10h, 1h and 10min splits to simulate varying amounts of labelled training data. 
Validation and test data are generated from LibriSpeech dev-clean and test-clean, respectively.

Binaural utterances are generated by sampling an utterance, a subject from the ARI HRTF dataset, and an RIR from the RIR dataset and an azimuth angle $\phi \sim \mathcal{U}(\SI{-90} {\degree}, \SI{90} {\degree})$ is sampled while fixing the elevation angle at \SI{90}{\degree} (horizontal plane). 
The mono speech signal is convolved with the RIR and the left-right HRTFs are then applied to generate the final binaural signal, which is saved as a .flac file.
The microphone array geometry and DoA are stored as metadata, enabling DoA prediction.
The static generated binaural dataset is used for supervised training/fine-tuning and benchmarking of models.
During pretraining binaural utterances are generated in an online fashion following the same procedure, using all \SI{960}{\hour} of speech from the librispeech training data as the basis.


\subsection{Pretraining Setup}

The general pretraining setup is shown in \Cref{fig:neural_gcc}. 
We pretrain a total of four models, each with a different spatial feature extractor, namely GCC, GCC-PHAT, CPS-PHAT phase and a lastly a the concatenation of ILD and IPD features\eqref{eq:cps_phat} as the pretraining target.
Pretraining uses the full \SI{960}{\hour} LibriSpeech training data, binauralized in an online fashion, as described in \Cref{subsec:binaural_dataset}, using the training HRTFs. 
Noise augmentation is done by adding diffuse environmental noise, generated with anf-generator\cite{anf_generator}, with an SNR between \SIrange{-20}{20}{\decibel}. 
Here, we use the same noise as in \cite{Kolbaek_noise_robust_speaker_verification} consisting of bus, babble, café, mixed, pedestrian, street and speech shaped noise \cite{Kolbaek_noise_robust_speaker_verification}.
We add noise at SNRs levels of \SIrange{-20}{20}{\decibel} in steps of \SI{5}{\decibel}.

We follow \cite{Yang_SSLSAR} and extract STFT features as in \eqref{eq:stft_real_imag}, using a window length of 400, a hop length of 160, and 512 FFT coefficients. 
The STFT features are passed to a 2-layer causal Conformer encoder with a 64-dimensional embedding, 4 attention heads, and a convolution kernel size of 31, similar to \cite{Yang_SSLSAR}, but with causal masking and a smaller embedding dimension for a more lightweight model. 
A final linear layer predicts the spatial feature targets, which are computed using the same STFT parameters as the encoder input.

We use an MSE loss, defined in \eqref{eq:mse_loss}, and train for 50k steps, using bucket batching with a bucket size of \SI{10}{\minute}, and save the best model based on validation performance. 
The learning rate follows a cosine annealing schedule with warm-up restarts, using a maximum learning rate of 1e-3, a minimum of 5e-7, 3k warm-up steps, and 30k steps per cycle. The cycle length scales by 0.9 and the maximum learning rate by 0.5 after each cycle. 
We use AdamW \cite{loshchilov_adamw} with $\beta=[0.9, 0.98]$ and $\varepsilon=1.e-6$, and a weight decay of 0.01. Pretraining takes approx. $\SI{14}{\hour}$ on a single NVIDIA A40 GPU.

\subsection{Supervised Fine-tuning}


For supervised fine-tuning, we use the same feature extractor and encoder as in pretraining, discarding the feature predictor and adding a linear classification layer for DoA prediction, as shown in \Cref{fig:neural_gcc}.
We predict DOAs with an angular resolution of \SI{5}{\degree}. 
During fine-tuning, we also apply diffuse noise augmentation as in pretraining.
The objective is a cross-entropy loss as described in \cref{subsec:doa_dnn}. 
The same hyperparameters as pretraining are used, as they were found to work well. 
However, we use a fixed batch size of 8 utterances and adjust the scheduler to have only 10 warm-up steps and 1k steps per cycle due to less training data. 
Fine-tuning on a single NVIDIA T4 GPU takes approx. \SI{4}{\hour} for the \SI{10}{\hour} training set.

To evaluate the fine-tuned model, we test on the binaural test-clean dataset under clean and noisy conditions. 
The noise types are the same as used during training; however, the individual noise clips are separate test clips different from the ones used for training. 
For all noise types we evaluate with SNRs between \SIrange{-20}{20}{\decibel} in steps of \SI{10}{\decibel}.

\subsection{Baselines}



We train three DNN-based reference systems using the same general setup as for fine-tuning, using either GCC, GCC-PHAT, or STFT features as the input.
This is followed by a 2-layer causal Conformer encoder and a linear classifier, same as our fine-tuned model.


Lastly, we evaluate a classic approach using argmax on GCC-PHAT features as in \eqref{eq:gcc_argmax_angle} applying average pooling using a window of 100 frames.
This yields a total of six reference baselines.
A summary of all DoA prediction setups is shown in \Cref{tab:model_table}.

\begin{table}[tb]
    \renewcommand{\arraystretch}{0.94}
    \setlength{\tabcolsep}{3pt}
    \centering
    \footnotesize
    \caption{Summary of evaluated models. SFD denotes pretraining with the proposed spatial feature distillation framework. 
    }
    \label{tab:model_table}
    \begin{tabular}{p{2.7cm} p{3.45cm} l}
    \toprule
    \textbf{Model} & \textbf{Architecture} & \textbf{\# Param.} \\
    \midrule
    GCCPHAT-Argmax & GCCPHAT-AvgPool-Argmax & 0 \\
    \midrule
    GCC-DNN & GCC-Conformer-Linear& 415.56 k \\
    GCCPHAT-DNN & GCCPHAT-Conformer-Linear& 415.56 k \\
    STFT-DNN & STFT-Conformer-Linear& 545 k \\
    \midrule
    SFD-GCC & STFT-Conformer-Linear & 545 k \\
    SFD-GCC-PHAT & STFT-Conformer-Linear & 545 k \\
    SFD-CPSPhase & STFT-Conformer-Linear & 545 k \\
    SFD-ILD+IPD & STFT-Conformer-Linear & 545 k \\
    \bottomrule
    \end{tabular}   
\end{table}

\section{Results}
In the experimental results analysis, we investigate the encoder output after pretraining, and whether the fine-tuned models outperform purely supervised DoA models.

\begin{figure*}[tb] 
    \centering
    \begin{minipage}{\textwidth}
        \renewcommand{\arraystretch}{0.94}
        \setlength{\tabcolsep}{3pt}
        \centering
        \footnotesize
        \captionof{table}{Mean angular error (MAE) test scores at different noise levels for models trained on 1h BinauralLibriLight when evaluated on the test set. Values are presented as $\text{mean} \pm \text{standard error}$. Standard errors were computed using the bootstrap method. SFD models are pretrained with the proposed framework.}
        \label{tab:performance_doa}
        \begin{tabular}{crrrrrrr}
            \toprule
            \multirow{2}{*}{Model} & \multicolumn{7}{c}{Noise level}\\
                  \cmidrule{2-8}
            & \multicolumn{1}{c}{\SI{-20}{\decibel}} & \multicolumn{1}{c}{\SI{-10}{\decibel}} & \multicolumn{1}{c}{\SI{0}{\decibel}} & \multicolumn{1}{c}{\SI{10}{\decibel}} & \multicolumn{1}{c}{\SI{20}{\decibel}} & \multicolumn{1}{c}{clean} & \multicolumn{1}{c}{Avg.}\\
            \midrule
            GCCPHAT-Argmax & $44.47 \pm 0.02$ & $41.02 \pm 0.02$ & $34.04 \pm 0.02$ & $27.30 \pm 0.02$ & $23.78 \pm 0.02$ & $17.44 \pm 0.02$ & $32.59 \pm 0.01$ \\
            \midrule
            GCC-DNN & $49.83 \pm 0.03$ & $19.01 \pm 0.02$ & $7.97 \pm 0.01$ & $5.96 \pm 0.01$ & $6.00 \pm 0.01$ & $5.97 \pm 0.01$ & $15.42 \pm 0.00$ \\
            GCCPHAT-DNN & $38.03 \pm 0.03$ & $17.44 \pm 0.02$ & $8.95 \pm 0.02$ & $5.94 \pm 0.01$ & $4.89 \pm 0.01$ & $4.29 \pm 0.01$ & $13.16 \pm 0.01$ \\
            STFT-DNN & $45.72 \pm 0.02$ & $24.71 \pm 0.01$ & $15.32 \pm 0.01$ & $13.96 \pm 0.01$ & $13.70 \pm 0.01$ & $13.58 \pm 0.01$ & $20.94 \pm 0.00$ \\
            \midrule
            SFD-GCC  & $34.19 \pm 0.02$ & $10.62 \pm 0.01$ & $6.39 \pm 0.01$ & $5.88 \pm 0.01$ & $5.83 \pm 0.01$ & $5.84 \pm 0.01$ & $10.92 \pm 0.00$ \\
            SFD-GCC-PHAT & $\mathbf{24.24} \pm 0.02$ & $\mathbf{7.00} \pm 0.01$ & $4.09 \pm 0.00$ & $3.81 \pm 0.00$ & $3.77 \pm 0.00$ & $3.77 \pm 0.00$ & $7.26 \pm 0.00$ \\
            SFD-CPSPhase & $24.53 \pm 0.02$ & $7.49 \pm 0.01$ & $\mathbf{3.62} \pm 0.00$ & $\mathbf{3.20} \pm 0.00$ & $\mathbf{3.16} \pm 0.00$ & $\mathbf{3.16} \pm 0.00$ & $\mathbf{7.05} \pm 0.00$ \\
            SFD-ILD+IPD  & $26.22 \pm 0.01$ & $10.94 \pm 0.01$ & $6.87 \pm 0.00$ & $6.61 \pm 0.00$ & $6.69 \pm 0.00$ & $6.71 \pm 0.00$ & $10.27 \pm 0.00$ \\
            \bottomrule
        \end{tabular}
    \end{minipage}
\end{figure*}

\begin{figure}
    \centering
    \includegraphics[width=\linewidth]{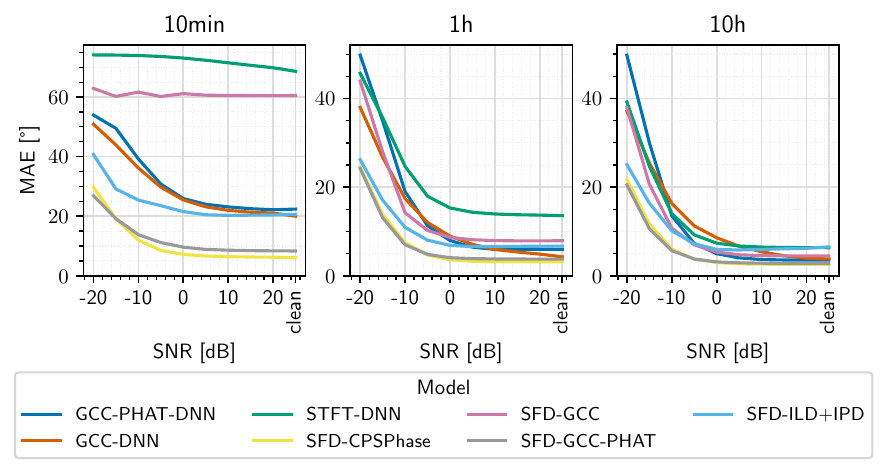}
    \caption{Comparison of mean angular error (MAE) at
    different $\mathrm{SNR}$ levels, varying the amount of training data.}
    \label{fig:model_accuracy}
\end{figure}



\subsection{DoA prediction performance}
\Cref{tab:performance_doa} compares the mean angular error (MAE) on the test set, averaged over all noise types, for models trained on the Binaural LibriLight 10h dataset. 
After fine-tuning for DoA prediction, all pretrained models show significant improvements in test set performance over reference methods. 
The SFD-CPSPhase model achieves the lowest MAE at all noise levels, outperforming the best supervised model (GCC-PHAT-DNN) by \SI{46.44}{\percent} on average. 
The SFD-GCC-PHAT model achieves a slightly worse average score than SFD-CPSPhase, however, it performs sligthly better for \SI{-10}{\decibel} and \SI{-20}{\decibel}.
SFD-GCC and ILD+IPD performs worst of the pretraining targets, suggesting that including ILD information in the pretraining target is detrimental to performance.

When varying the amount of labelled data, \Cref{fig:model_accuracy} shows that pretraining with GCC-PHAT and CPS Phase consistently yields good performance, even with as little as \SI{10}{\minute} of labelled data. 
In contrast, the supervised DNN baselines suffer significantly from data limitations. 
This underscores the value of our proposed framework in low-data settings for learning robust spatial representations without DoA labels, similar to the use case of SSL features in other speech processing applications \cite{baevski_wav2vec2}.

\section{Conclusion}
We presented a framework method for learning a robust spatial representation from unlabelled binaural audio data by predicting spatial features. 
Evaluations in different noisy and reverberant environments show that it outperforms both classic and supervised approaches for DoA estimation of a single static speaker, reducing the mean angular error by \SI{33.6}{\percent} on average. 
A limitation of our current work, is the assumption of clean speech being available.
Future work will explore scenarios with no available clean speech, as well as handling of moving sources, multiple speakers, and the effects of scaling pretraining data and model size.

\clearpage
\bibliographystyle{IEEEtran}
\bibliography{refs25}

\begin{thebibliography}{10}
\providecommand{\url}[1]{#1}
\csname url@samestyle\endcsname
\providecommand{\newblock}{\relax}
\providecommand{\bibinfo}[2]{#2}
\providecommand{\BIBentrySTDinterwordspacing}{\spaceskip=0pt\relax}
\providecommand{\BIBentryALTinterwordstretchfactor}{4}
\providecommand{\BIBentryALTinterwordspacing}{\spaceskip=\fontdimen2\font plus
\BIBentryALTinterwordstretchfactor\fontdimen3\font minus \fontdimen4\font\relax}
\providecommand{\BIBforeignlanguage}[2]{{%
\expandafter\ifx\csname l@#1\endcsname\relax
\typeout{** WARNING: IEEEtran.bst: No hyphenation pattern has been}%
\typeout{** loaded for the language `#1'. Using the pattern for}%
\typeout{** the default language instead.}%
\else
\language=\csname l@#1\endcsname
\fi
#2}}
\providecommand{\BIBdecl}{\relax}
\BIBdecl

\bibitem{Grothe_mammal_hearing}
B.~Grothe, M.~Pecka, and D.~McAlpine, ``Mechanisms of sound localization in mammals,'' \emph{Physiological Reviews}, vol.~90, no.~3, pp. 983--1012, 2010.

\bibitem{Knapp_GCC_for_time_delay_estimation}
C.~Knapp and G.~Carter, ``The generalized correlation method for estimation of time delay,'' \emph{IEEE Transactions on Acoustics, Speech, and Signal Processing}, vol.~24, no.~4, pp. 320--327, 1976.

\bibitem{Omologo_CPS_event_location}
M.~Omologo and P.~Svaizer, ``Use of the crosspower-spectrum phase in acoustic event location,'' \emph{IEEE Transactions on Speech and Audio Processing}, vol.~5, no.~3, pp. 288--292, 1997.

\bibitem{Benesty_microphone_array_SP}
J.~Benesty, J.~Cheng, and Y.~Huang, \emph{Microphone Array Signal Processing}, ser. Springer Topics in Signal Processing.\hskip 1em plus 0.5em minus 0.4em\relax Springer, 2008, vol.~1.

\bibitem{Blandin_GCCPHAT_clustering}
C.~Blandin, A.~Ozerov, and E.~Vincent, ``Multi-source tdoa estimation in reverberant audio using angular spectra and clustering,'' \emph{Signal Processing}, vol.~92, no.~8, pp. 1950--1960, 2012.

\bibitem{Xiao_GCC_learning_based_DoA}
X.~Xiao, S.~Zhao, X.~Zhong, D.~L. Jones, E.~S. Chng, and H.~Li, ``A learning-based approach to direction of arrival estimation in noisy and reverberant environments,'' in \emph{2015 IEEE International Conference on Acoustics, Speech and Signal Processing (ICASSP)}, 2015, pp. 2814--2818.

\bibitem{Grinstein_SRP_tutorial}
E.~Grinstein, E.~Tengan, B.~Çakmak, T.~Dietzen, L.~Nunes, T.~Waterschoot, M.~Brookes, and P.~Naylor, ``Steered response power for sound source localization: a tutorial review,'' \emph{EURASIP Journal on Audio, Speech, and Music Processing}, vol. 2024, 11 2024.

\bibitem{Schmidt_MUSIC}
R.~Schmidt, ``Multiple emitter location and signal parameter estimation,'' \emph{IEEE Transactions on Antennas and Propagation}, vol.~34, no.~3, pp. 276--280, 1986.

\bibitem{Fuchs_monaural_sound_localization}
A.~K. Fuchs, C.~Feldbauer, and M.~Stark, ``Monaural sound localization,'' in \emph{Proc. Interspeech}, 2011, pp. 2521--2524.

\bibitem{Ma_DNN_robust_localization}
N.~Ma, T.~May, and G.~J. Brown, ``Exploiting deep neural networks and head movements for robust binaural localization of multiple sources in reverberant environments,'' \emph{IEEE/ACM Transactions on Audio, Speech, and Language Processing}, vol.~25, no.~12, pp. 2444--2453, 2017.

\bibitem{yalta_dnn_doa}
N.~Yalta, K.~Nakadai, and T.~Ogata, ``Sound source localization using deep learning models,'' \emph{Journal of Robotics and Mechatronics}, vol.~29, no.~1, pp. 37--48, 2017.

\bibitem{subramanian_dnn_multisource_localization}
A.~S. Subramanian, C.~Weng, S.~Watanabe, M.~Yu, and D.~Yu, ``Deep learning based multi-source localization with source splitting and its effectiveness in multi-talker speech recognition,'' \emph{Computer Speech \& Language}, vol.~75, p. 101360, 2022.

\bibitem{Abdelrahman_ssl_speech_review}
A.~Mohamed, H.-y. Lee, L.~Borgholt, J.~D. Havtorn, J.~Edin, C.~Igel, K.~Kirchhoff, S.-W. Li, K.~Livescu, L.~Maaløe, T.~N. Sainath, and S.~Watanabe, ``Self-supervised speech representation learning: A review,'' \emph{IEEE Journal of Selected Topics in Signal Processing}, vol.~16, no.~6, pp. 1179--1210, 2022.

\bibitem{Pak_phase_difference_enhancement}
J.~Pak and J.~W. Shin, ``Sound localization based on phase difference enhancement using deep neural networks,'' \emph{IEEE/ACM Transactions on Audio, Speech, and Language Processing}, vol.~27, no.~8, pp. 1335--1345, 2019.

\bibitem{Cheng_steering_vector_enhancement}
L.~Cheng, X.~Sun, D.~Yao, J.~Li, and Y.~Yan, ``Estimation reliability function assisted sound source localization with enhanced steering vector phase difference,'' \emph{IEEE/ACM Transactions on Audio, Speech, and Language Processing}, vol.~29, pp. 421--435, 2021.

\bibitem{Wang_IPDNet}
Y.~Wang, B.~Yang, and X.~Li, ``Ipdnet: A universal direct-path ipd estimation network for sound source localization,'' \emph{IEEE/ACM Transactions on Audio, Speech, and Language Processing}, vol.~32, pp. 5051--5064, 2024.

\bibitem{Huang_masked_audio_MAE}
P.-Y. Huang, H.~Xu, J.~Li, A.~Baevski, M.~Auli, W.~Galuba, F.~Metze, and C.~Feichtenhofer, ``Masked autoencoders that listen,'' in \emph{Proc. NeurIPS}, vol.~35, 2022, pp. 28\,708--28\,720.

\bibitem{unsupervised_AR_model_for_speech}
Y.~Chung, W.~Hsu, H.~Tang, and J.~R. Glass, ``An unsupervised autoregressive model for speech representation learning,'' in \emph{Proc. Interspeech}, 2019, pp. 146--150.

\bibitem{baevski_wav2vec2}
A.~Baevski, Y.~Zhou, A.~Mohamed, and M.~Auli, ``wav2vec 2.0: A framework for self-supervised learning of speech representations,'' in \emph{Proc. NeurIPS}, vol.~33, 2020, pp. 12\,449--12\,460.

\bibitem{HuBERT_paper}
W.-N. Hsu, B.~Bolte, Y.-H.~H. Tsai, K.~Lakhotia, R.~Salakhutdinov, and A.~rahman Mohamed, ``{HuBERT}: Self-supervised speech representation learning by masked prediction of hidden units,'' \emph{IEEE/ACM Transactions on Audio, Speech, and Language Processing}, vol.~29, pp. 3451--3460, 2021.

\bibitem{Chen_WavLMLS}
S.~Chen, C.~Wang, Z.~Chen, Y.~Wu \emph{et~al.}, ``{WavLM}: Large-scale self-supervised pre-training for full stack speech processing,'' \emph{IEEE Journal of Selected Topics in Signal Processing}, vol.~16, pp. 1505--1518, 2021.

\bibitem{Chen_BEATS_SSL}
S.~Chen, Y.~Wu, C.~Wang, S.~Liu, D.~Tompkins, Z.~Chen, W.~Che, X.~Yu, and F.~Wei, ``Beats: audio pre-training with acoustic tokenizers,'' in \emph{Proceedings of the 40th International Conference on Machine Learning}.\hskip 1em plus 0.5em minus 0.4em\relax JMLR.org, 2023.

\bibitem{oord_cpc}
A.~van~den Oord, Y.~Li, and O.~Vinyals, ``Representation learning with contrastive predictive coding,'' \emph{ArXiv}, vol. abs/1807.03748, 2018.

\bibitem{Chen_XEUS}
W.~Chen, W.~Zhang, Y.~Peng, X.~Li, J.~Tian, J.~Shi, X.~Chang, S.~Maiti, K.~Livescu, and S.~Watanabe, ``Towards robust speech representation learning for thousands of languages,'' in \emph{Proceedings of the 2024 Conference on Empirical Methods in Natural Language Processing}.\hskip 1em plus 0.5em minus 0.4em\relax Association for Computational Linguistics, Nov. 2024, pp. 10\,205--10\,224.

\bibitem{Zhu_noise_robust_ssl_asr}
Q.-S. Zhu, J.~Zhang, Z.-Q. Zhang, M.-H. Wu, X.~Fang, and L.-R. Dai, ``A noise-robust self-supervised pre-training model based speech representation learning for automatic speech recognition,'' in \emph{Proc. ICASSP}, 2022, pp. 3174--3178.

\bibitem{Munakata_joint_sep_and_loc_FSC}
H.~Munakata, Y.~Bando, R.~Takeda, K.~Komatani, and M.~Onishi, ``Joint separation and localization of moving sound sources based on neural full-rank spatial covariance analysis,'' \emph{IEEE Signal Processing Letters}, vol.~30, pp. 384--388, 2023.

\bibitem{Wang_UNSSOR}
Z.-Q. Wang and S.~Watanabe, ``Unssor: Unsupervised neural speech separation by leveraging over-determined training mixtures,'' in \emph{Proc. NeurIPS}, vol.~36, 2023, pp. 34\,021--34\,042.

\bibitem{Yang_SSLSAR}
B.~Yang and X.~Li, ``Self-supervised learning of spatial acoustic representation with cross-channel signal reconstruction and multi-channel conformer,'' \emph{IEEE/ACM Transactions on Audio, Speech, and Language Processing}, vol.~32, pp. 4211--4225, 2024.

\bibitem{librispeech}
V.~Panayotov, G.~Chen, D.~Povey, and S.~Khudanpur, ``Librispeech: An {ASR} corpus based on public domain audio books,'' in \emph{Proc. ICASSP}, 2015, pp. 5206--5210.

\bibitem{librilight}
J.~{Kahn}, M.~{Rivière}, W.~{Zheng}, E.~{Kharitonov}, Q.~{Xu}, P.~E. {Mazaré}, J.~{Karadayi}, V.~{Liptchinsky}, R.~{Collobert}, C.~{Fuegen}, T.~{Likhomanenko}, G.~{Synnaeve}, A.~{Joulin}, A.~{Mohamed}, and E.~{Dupoux}, ``{Libri-Light}: A benchmark for asr with limited or no supervision,'' in \emph{Proc. ICASSP}, 2020, pp. 7669--7673.

\bibitem{ari_hrtf}
{Institut für Schallforschung der Österreichischen Akademie der Wissenschaften}, ``{HRTF-DATABASE},'' \url{https://www.oeaw.ac.at/isf/das-institut/software/hrtf-database}, 2024.

\bibitem{Kowalk_DNN_signal_informed_DoA_w_GCCPHAT}
U.~Kowalk, S.~Doclo, and J.~Bitzer, ``Signal-informed dnn-based doa estimation combining an external microphone and gcc-phat features,'' in \emph{{Proc. IWAENC}}, 2022, pp. 1--5.

\bibitem{Kowalk_DoA_gemoetry_aware_dnn}
------, ``Geometry-aware doa estimation using a deep neural network with mixed-data input features,'' in \emph{Proc. ICASSP}, 2023, pp. 1--5.

\bibitem{Goli_STFT_DNN_localization}
P.~Goli and S.~van~de Par, ``Deep learning-based speech specific source localization by using binaural and monaural microphone arrays in hearing aids,'' \emph{IEEE/ACM Transactions on Audio, Speech, and Language Processing}, vol.~31, pp. 1652--1666, 2023.

\bibitem{Tang_RegressionAC}
Z.~Tang, J.~D. Kanu, K.~Hogan, and D.~Manocha, ``Regression and classification for direction-of-arrival estimation with convolutional recurrent neural networks,'' in \emph{Interspeech}, 2019.

\bibitem{Gulati_conformer_interspeech}
A.~Gulati, J.~Qin, C.-C. Chiu, N.~Parmar, Y.~Zhang, J.~Yu, W.~Han, S.~Wang, Z.~Zhang, Y.~Wu, and R.~Pang, ``Conformer: Convolution-augmented transformer for speech recognition,'' in \emph{Proc. Interspeech}, {Shanghai, China}, 10 2020, pp. 5036--5040.

\bibitem{Barker_Clarity_challenge}
J.~Barker, M.~Akeroyd, T.~J. Cox, J.~F. Culling, J.~Firth, S.~Graetzer, H.~Griffiths, L.~Harris, G.~Naylor, Z.~Podwinska \emph{et~al.}, ``The 1st clarity prediction challenge: A machine learning challenge for hearing aid intelligibility prediction.'' in \emph{Proc. Interspeech}, 2022, pp. 3508--3512.

\bibitem{Ko_RIR_data_augmentation}
T.~Ko, V.~Peddinti, D.~Povey, M.~L. Seltzer, and S.~Khudanpur, ``A study on data augmentation of reverberant speech for robust speech recognition,'' in \emph{Proc. ICASSP}, 2017, pp. 5220--5224.

\bibitem{anf_generator}
{International Audio Laboratories Erlangen}, ``{anf-generator},'' \url{https://github.com/audiolabs/anf-generator}, 2025.

\bibitem{Kolbaek_noise_robust_speaker_verification}
M.~Kolbœk, Z.-H. Tan, and J.~Jensen, ``Speech enhancement using long short-term memory based recurrent neural networks for noise robust speaker verification,'' in \emph{2016 IEEE Spoken Language Technology Workshop (SLT)}, 2016, pp. 305--311.

\bibitem{loshchilov_adamw}
I.~Loshchilov and F.~Hutter, ``Decoupled weight decay regularization,'' in \emph{International Conference on Learning Representations}, 2019.

\end{thebibliography}







\end{document}